\documentclass{pazh_engl}
\usepackage{graphicx,natbib}
\bibpunct{(}{)}{;}{a}{}{,}

\begin{document}


\title{\Large OPTICAL IDENTIFICATION OF A NEW CATACLYSMIC VARIABLE FROM INTEGRAL ALL SKY SURVEY: IGR J08390--4833}

   \author{\copyright 2009 \bfseries M.~Revnivtsev\inst{1,2,3}, A. Kniazev \inst{4,5}, S.~Sazonov 
\inst{1,2}, R.~Burenin \inst{2}, \\ A.~Tekola \inst{4}, D.A.H.~Buckley \inst{4}, 
M.L.~Pretorius \inst{4}, J.~Menzies\inst{4}, W.~Lawson \inst{6}}

   \institute{
              Max-Planck-Institute f\"ur Astrophysik,
              Karl-Schwarzschild-Str. 1, D-85740 Garching bei M\"unchen,
              Germany,
     \and
              Space Research Institute, Russian Academy of Sciences,
              Profsoyuznaya 84/32, 117997 Moscow, Russia
     \and
	   Excellence Cluster Universe, Technische Universität München, Boltzmannstrasse 2, D-85740  Garching bei M\"unchen, Germany
	\and
 South African Astronomical Observatory, Observatory Rd, Observatory 7925, South Africa 
 \and
 Special Astrophysical Observatory, Nizhnij Arkhyz, Karachai-Circassia,
369167, Russia
\and
School of Physical, Environmental and Mathematical Sciences, University
of New South Wales, Australian Defence Force Academy, Canberra ACT 2600,
Australia 
 }
\authorrunning{M.~Revnivtsev et al.}
        \titlerunning{Optical identification of IGR J08390--4833}

\abstract{We have optically identified a recently
discovered INTEGRAL source, IGR J08390--4833, with a cataclysmic variable, i.e. an accreting white dwarf in a binary
system. The spectrum exhibits a rising blue continuum together with Balmer and HeII emission lines.
Analysis of the light curve of the source shows clear presence of intrinsic variability on a time scale
of the order of an hour, although we do not claim that this variability is
periodic. Therefore we are not yet able to classify the object into a specific CV subclass.
  \keywords{cataclysmic variables --- X-ray sources --- optical
    observations}
}
\maketitle

\section*{\sf INTRODUCTION}
Surveys of the sky in X-ray energy band are very useful in constructing
catalogs of different classes of sources with minimal biases. 
These catalogs, in turn, 
provides us a possibility to study properties of populations of
different astrophysical objects. However, in order for such measurements to be
accurate one needs to minimize incompleteness of the catalogs.

INTEGRAL observatory \citep{winkler03} is performing the most
sensitive all sky survey in hard X-ray energy band up to date 
\citep{krivonos07} and provides important information about
populations of nearby AGNs \citep{sazonov07}, X-ray binaries 
\citep{lutovinov05,bodaghee07,revnivtsev08a} and cataclysmic variables 
\citep{revnivtsev08b}. 

The catalog contains more than 400 sources, many of which
still lack a secure optical identification. Our group systematically performs
optical observations of unidentified INTEGRAL sources with the aim
of determining their nature \citep{bikmaev06a,bikmaev06b,burenin06a,burenin06b,mescheryakov06,bikmaev08,burenin08}. 
The increase of the completeness of the
INTEGRAL all sky catalog significantly increases its scientific value.
In the present paper we report on the identification of the
source IGR J08390--4833 as a new cataclysmic variable.

\section*{\sf OBSERVATIONS}

\begin{figure}[htb]
\centerline{
\includegraphics[width=0.8\columnwidth]{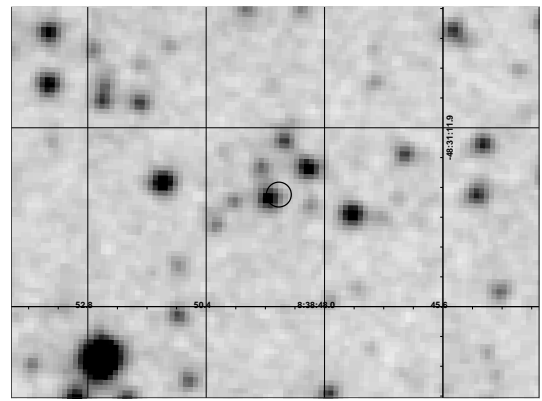}
}
\caption{A 1.4x1.2 arcmin region of the sky around IGR J08390--4833 from the DSS2-R all-sky survey.
The position of the source from Chandra observations is shown 
by a circle where the radius denotes the X sigma uncertainty.  
}
\label{image}
\end{figure}

The source IGR J08390--4833 was discovered in INTEGRAL observations by
\cite{sazonov08}and follow-up observations 
with Chandra \citep{sazonov08} led to a likely optical identification with a V $\sim 16$ magnitude star 
at RA=08h 38m 49.11s, Dec=-48d 31m 24.7s. An image of a [1.4x1.2 arcmin] region
around the object from the DSS2-R all sky survey is shown in  Fig.~\ref{image}.

\subsection*{\sf SPECTRAL}

Our spectroscopic observations were carried out with the
SAAO 1.9m telescope during three nights in April 2008.
The observations were performed with the Cassegrain spectrograph
using a long slit of 3$^{\prime} \times$1.5$^{\prime\prime}$. A grating 
with 300 lines mm$^{-1}$ was used
with the SITe 266$\times$1798 pixel CCD detector.
This gave a wavelength range $\sim$3500--7300~\AA\
with $\sim$2.3 \AA~pixel$^{-1}$ and FWHM $\sim$7~\AA\
along the dispersion direction and
the scale along the slit 0\farcs7 pixel$^{-1}$.
The seeing during the observations varied from
1.0$^{\prime\prime}$ to 1.6$^{\prime\prime}$,
but was stable during each night. Spectra of Cu--Ar comparison arcs were obtained to calibrate the wavelength
scale. 
The airmass range was from 1.04 to 1.18 and
spectrophotometric flux standards were observed
after object for flux calibration.

The reduction of all data was performed using the IRAF\footnote[1]{http://iraf.noao.edu/} data
reduction package. Cosmic ray hits were
removed from the 2D spectral using MIDAS. We corrected for the overscan, subtracted the bias
and performed flat-field corrections ussing IRAF tasks in the
{\em ccdred}\ package and used the software tasks in {\em twodspec} to perform
the wavelength calibration and to correct each frame for distortion and tilt. The accuracy of the wavelength calibration
was better than 0.4 \AA.
After the 2D spectrum was wavelength-calibrated,
the night sky background was subtracted.  Using our data of the
spectrophotometry standard stars, the intensities of the 2D spectrum
were transformed to absolute fluxes.  One-dimensional spectra were
extracted in order to get the total flux.

\subsection*{\sf IMAGING AND PHOTOMETRY}

On 26 April 2008 we obtained a light curve of the source
with using the SAAO 1m telescope and UCT CCD. The aim of the observation was 
to study its variability and to search for possible periodicities,caused by either by rotational or orbital modulations.

The high-speed photometry was taken with the University of Cape 
Town CCD photometer (UCT CCD, \cite{odonoghue95}) on the SAAO 1-m telescope. 
We used 10-s exposures (there is no dead time between exposures, 
since the photometer is a frame transfer CCD) with no filter. Unfiltered observations with the UCT CCD 
gives photometry with an effective wavelength similar to Johnson $V$-band, but with a very broad bandpass.  
The use of white light means that the observations cannot be precisely 
placed on a standard photometric system; the magnitude calibration 
approximates Johnson V only to within $\simeq$ 0.1 mag.  We performed 
differential photometry, implying that colour differences between the 
targets and comparison stars were ignored in correcting the data for 
atmospheric extinction.

Due to the detection of nebular emission lines around the 
object (see below), some H$\alpha$
images were obtained with the SAAO 1.0-m telescope.
One pair of images was obtained in V and H$\alpha$ on the night of 3 May 2008 while
four pairs of images were obtained in R and H$\alpha$ on the night of 
10 May 2008, with exposure times of 300 and 1200 sec, respectively
using the SITe 512$\times$512 pixel CCD camera (scale 0.31 arcsec/pixel).
The data reduction was performed using MIDAS and IRAF data
reduction packages. Cosmic rays were
removed from the 2D spectral frame using MIDAS.  IRAF tasks in the
{\em ccdred}\ package were used to correct for the overscan
and to perform flat-field corrections.

\section*{\sf RESULTS}

\subsection*{\sf SPECTRA}

\begin{figure}[htb]
\includegraphics[width=\columnwidth,bb=41 180 580 710,clip]{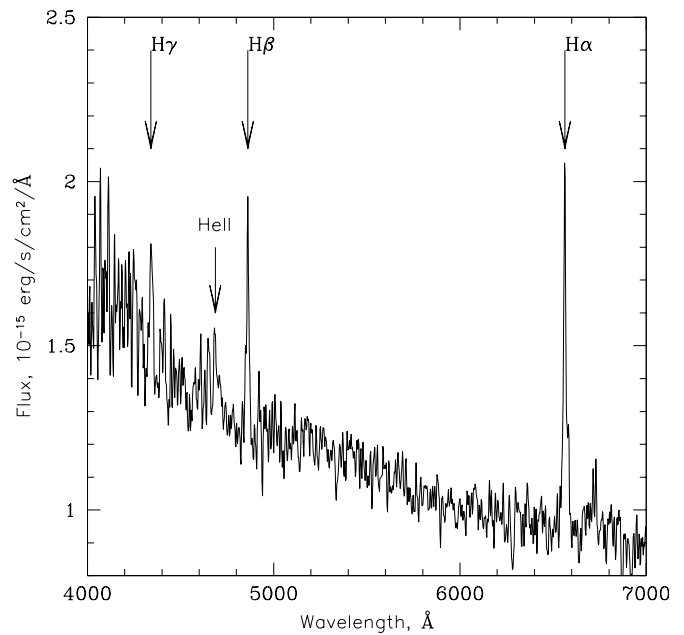}
\caption{Spectrum of IGRJ08390--4833, averaged over all observations}
\label{spectrum}
\end{figure}

\begin{figure}[htb]
\includegraphics[width=\columnwidth]{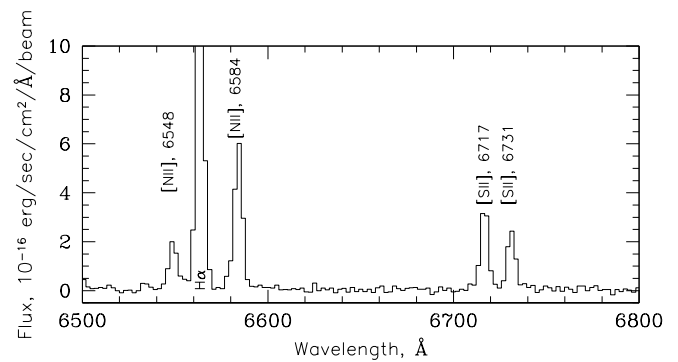}
\caption{Part of the spectrum of the nebula detected around IGRJ08390--4833}
\label{nebula_sp}
\end{figure}

\begin{figure}[htb]
\begin{center}
\vbox{
\includegraphics[width=0.8\columnwidth,bb=43 184 567 607,clip]{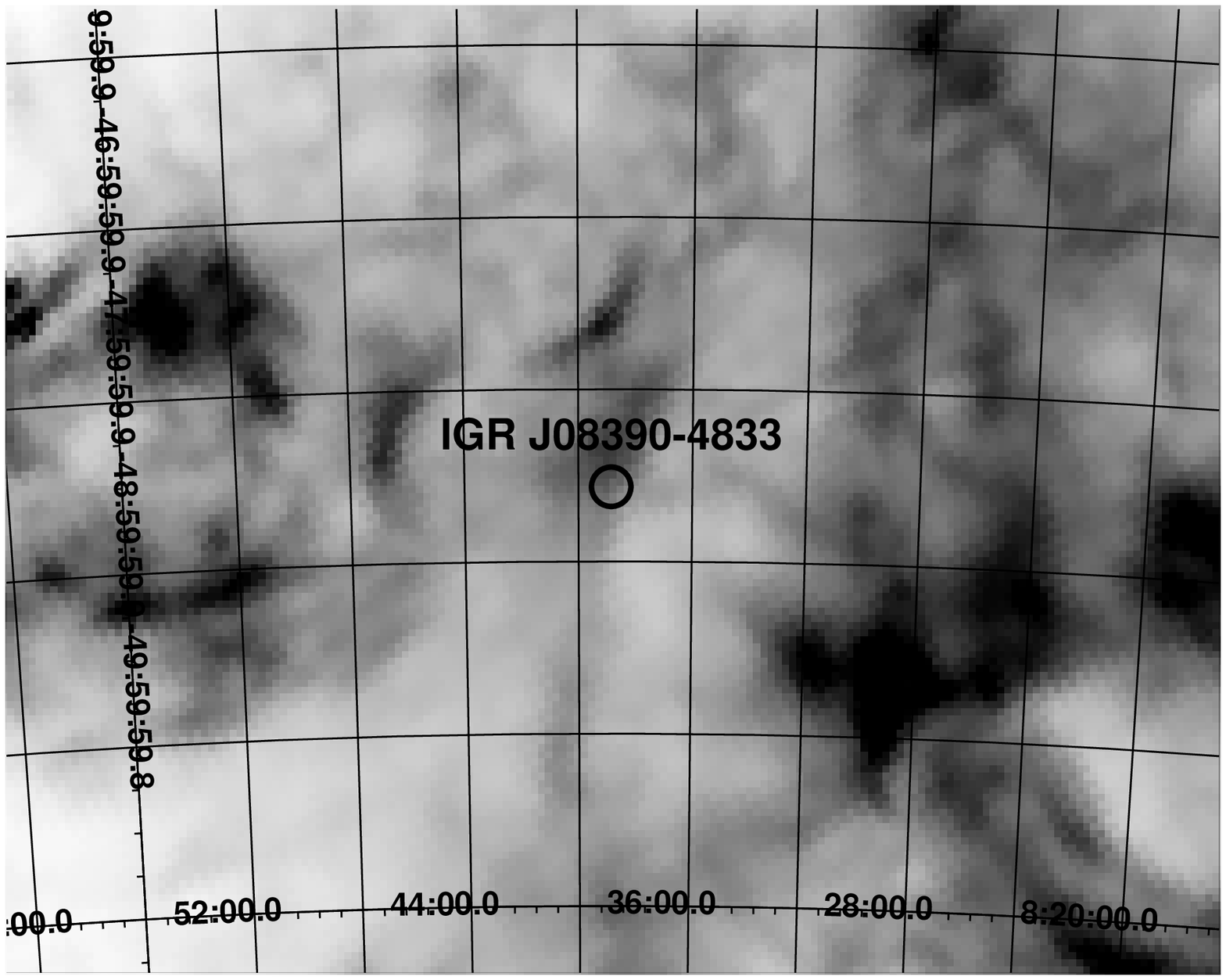}
\includegraphics[width=0.8\columnwidth,bb=23 3 500 462,clip]{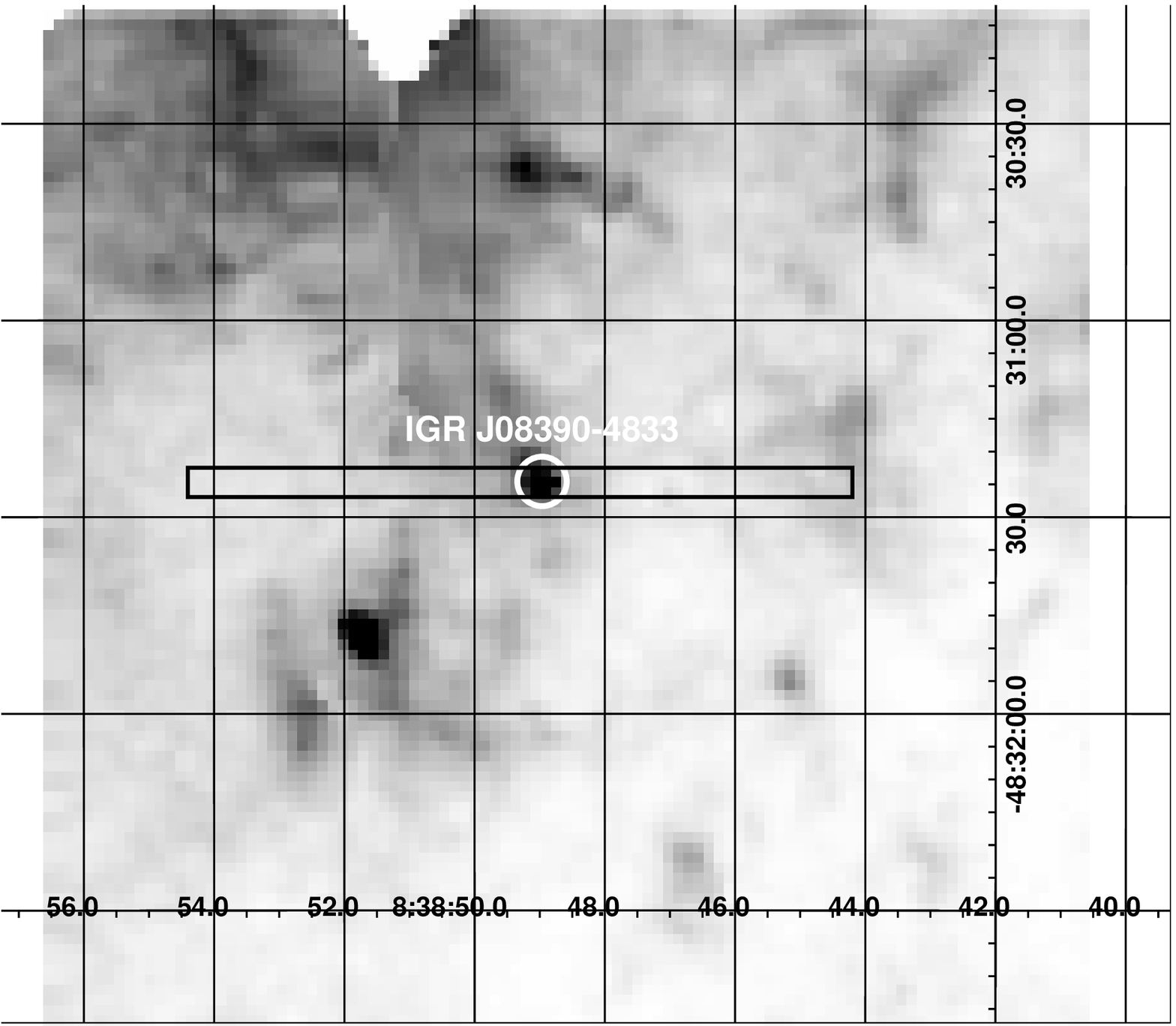}
}
\end{center}
\caption{{\sl Upper panel}: Image of a 7.5x6.1 degrees region around IGR J08390-4833 taken with a
narrow (bandpass $\sim$17\AA) $H\alpha$ filter (from Skyview service of HEASARC, image based in 
\citealt{finkbeiner03}). The position of IGR J08390-4833 is marked by a circle.
{\sl Lower panel}: A smaller image (2.6x2.4 arcmin) around IGR J08390-4833 taken with an $H\alpha$ filter, 
observed with the SAAO 1m telescope. Stars and other artefacts were
removed from the image. The bright $H\alpha$ image of IGR J08390-4833
is marked by a circle. The rectangular box denotes the size and position of the spectrograph slit used
to obtain the spectrum of IGR J08390-4833 and the nebula.}
\label{halphaimage}
\end{figure}

The grand-sum of all the spectra of IGR J08390--4833 obtained with SAAO 1.9m telescope 
is shown in Fig.~\ref{spectrum}. The bright Balmer emission lines with 
zero redshift immidiately show that the source is Galactic.
The shape of the continuumoptical spectrum along with a sequence of strong Balmer
emission lines, indicates that this source is a cataclysmic
variable \cite[see e.g.][]{schmidt86,bikmaev06b,masetti07}.

\subsection*{\sf EMISSION LINE NEBULA}

The 2D spectrum obtained by the long slit of the spectrograph indicates spatially extended emission lines, arising from nebulosity surrounding the source,
including  OII, [NII] 6548, 6584\AA\  and [S II]6716, 6731\AA\. The presence of these lines in the source spectrum is most likely a result of
imperfect subtraction of the bright nebula emission lines, which are clearly visible in 2D spectra.

It is interesting to check the possible casual connection of the
cataclysmic variable with this nebula, because cataclysmic
variables sometime might undergone classical nova explosion and this
might create some remnant/nebula \cite[see e.g.][]{slavin95}.
For this purpose we analized the spectrum of the nebula and
obtained $H\alpha$ image of the area.

The spectrum of the nebula is shown in Fig.\ref{nebula_sp}. Measured 
line ratios are listed in Table \ref{neb_lines}.
All emission lines were measured applying the MIDAS programs described in
detail in \cite{kniazev04}.
All overlapping lines were fitted
simultaneously as a blend of two or more Gaussian features:
the H$\alpha$ $\lambda$6563 and [\ion{N}{ii}] $\lambda\lambda$6548,6584
lines and the [\ion{S}{ii}] $\lambda\lambda$6716,6731 lines.
Parameters of the spectrum was analyzed using diagnostics of
\cite{kniazev08a}, which gave us values of the 
reddening in the spectrum and the density of emitting plasma.
For the latter we assumed  all lines are produced in an isothermal gas 
at uniform density and ionization level. The reddening correction,
electron temperatures and density were calculated iteratively
until the values converged.

\begin{table}
\caption{Ratios of line intensities (measured $F$ and corrected for 
interstellar reddening $I$) of the nebula around IGR J08390--4833. Also 
shown are: $EW(H\beta)$ -- equivalent width of the $H\beta$ line, C(H$\beta$) -- extinction coefficient, $A_B$ -- spectral reddening and $N_e$ -- density
of emitting plasma
}
\begin{tabular}{lcc} \hline
\rule{0pt}{10pt}
$\lambda_{0}$(\AA) Ion                  & F($\lambda$)/F(H$\beta$)&I($\lambda$)/I(H$\beta$) \\ \hline
3727\ [O\ {\sc ii}]\                           & 3.69$\pm$0.29 & 3.90$\pm$0.32 \\
4101\ H$\delta$\                               & 0.29$\pm$0.07 & 0.30$\pm$0.10 \\
4340\ H$\gamma$\                               & 0.48$\pm$0.07 & 0.50$\pm$0.08 \\
4861\ H$\beta$\                                & 1.00$\pm$0.08 & 1.00$\pm$0.09 \\
4959\ [O\ {\sc iii}]\                          & 0.17$\pm$0.06 & 0.17$\pm$0.06 \\
5007\ [O\ {\sc iii}]\                          & 0.45$\pm$0.05 & 0.45$\pm$0.05 \\
5869\ He\ {\sc ii}\                            & 0.09$\pm$0.04 & 0.09$\pm$0.04 \\
6548\ [N\ {\sc ii}]\                           & 0.50$\pm$0.05 & 0.47$\pm$0.04 \\
6563\ H$\alpha$\                               & 3.13$\pm$0.21 & 2.94$\pm$0.22 \\
6584\ [N\ {\sc ii}]\                           & 1.47$\pm$0.10 & 1.39$\pm$0.10 \\
6717\ [S\ {\sc ii}]\                           & 0.83$\pm$0.06 & 0.78$\pm$0.06 \\
6731\ [S\ {\sc ii}]\                           & 0.59$\pm$0.04 & 0.55$\pm$0.04 \\
& & \\
EW(H$\beta$)\ \AA\        & \multicolumn {2}{c}{ 131$^a$}   \\   
& & \\
C(H$\beta$)\ dex          & \multicolumn {2}{c}{0.08} \\
A$_B$\ mag                & \multicolumn {2}{c}{0.23} \\
& & \\
$[$\ion{S}{ii}$]$~$\lambda$6731/$\lambda$6717         & \multicolumn {2}{c}{1.409$\pm$0.156}   \\
$N_{\rm e}$(\ion{S}{ii} $\lambda$6731/$\lambda$6717), cm$^{-3}$  & \multicolumn{2}{c}{20$_{-10}^{+135}$} \\
\hline
\end{tabular}
\begin{list}{}
\item $^a$ -- Due to extreme weakness of the nebular continuum we can not securely
determine the uncertainty of the value of the equivalent width. However
we anticipate that it does not exceed 30\%.
\end{list}
\label{neb_lines}
\end{table}

The lines intensity ratios indicate 
that they originate in a hot low density region, which is likely
to be an extended HII region \cite[see e.g. disgnostics of HII region 
in][]{kniazev08b} which is clearly visible 
on the $H\alpha$ images \citep[see also 
Fig.\ref{halphaimage}]{finkbeiner03}. 

In order to check the association of the emission line nebula with
our source we have obtained an image of the sky around the source with an
 $H\alpha$ filter. The resulted ($H\alpha-$continum) colour image
was cleaned to remove residuals left from imperfect subtraction of
stars, and then binned into $1.5^{\prime\prime}\times1.5^{\prime\prime}$ pixels
(see Fig.~\ref{halphaimage}). The image does not show
a clear association of the nebula with the source.

\subsection*{\sf LIGHT CURVE}

\begin{figure}[htb]
\includegraphics[width=\columnwidth,bb=41 150 580 700,clip]{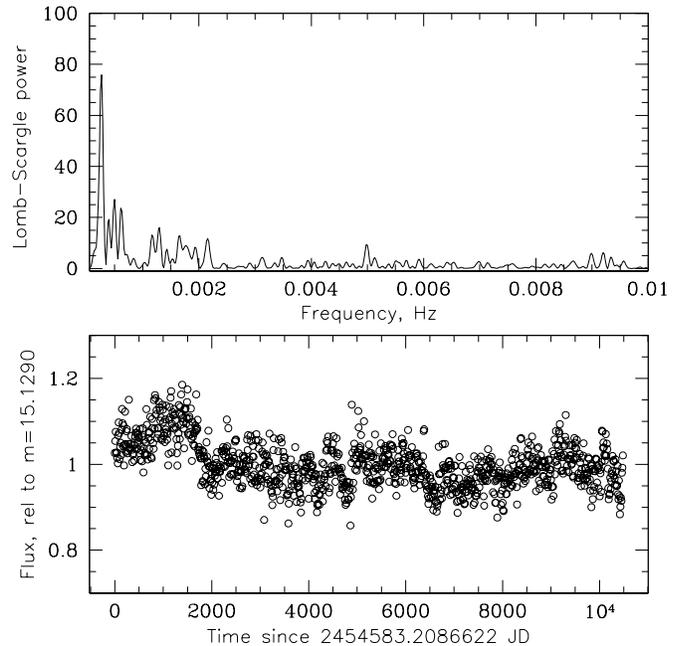}
\caption{Lower panel: light curve of IGR J08390-4833. Upper panel: 
Lomb-Scarge periodogram of de-trended light curve of the source.
}
\label{lcurve}
\end{figure}

The optical light curve of the source is presented in 
Fig.~\ref{lcurve} ({\sl lower panel}).
The Lomb-Scargle periodogram \citep{lomb76} is presented in 
Fig.~\ref{lcurve}({\sl upper panel}). Before construction of the Lomb-Scarge
periodogram the light curve was de-trended by subtraction of the best fit
quadratic function from the original data.

In absence of internal variability of the source flux the probability 
for Lomb-Scargle power to exceed the value $x$ is $P(>x)=\exp(-x)$.
Thus we can securely claim the presence of the instrinsic variability 
of the source flux at Fourier frequencies $f<7\times10^{-4}$ Hz.
The highest peak in the periodogram corresponds to the frequency
 $(2.6\pm0.3)\times 10^{-4}$ Hz, or to the period $P\sim 1.1\pm0.1$ hour.

Taking into account that the length of our time series
is only $\sim$3 hours, we cannot claim detection of any periodic variations. 
In order to make secure conclusion in this regard we need to obtain more observations 
spanning a longer time base.

\bigskip

Further classification of the source to subclasses of cataclysmic variables
can be only done after collection of more data in optical and X-ray energy
bands.

\begin{acknowledgements}

This work was supported by grants RFFI~07-02-01051,
07-02-00961, NSH-1789-2003.2, NSH-1100.2006.2, and the programm of
Presidium of Russian Adademy of Sciences ``Formation and evolution of
stars and galaxies''

\end{acknowledgements}

\end{document}